\documentclass[titlepage,12pt,twoside]{article}
\usepackage{amssymb,epsfig,pslatex}
\usepackage{cite}  
\usepackage{float}       
\usepackage{wrapfig}  
\usepackage{hhline} 
\usepackage{dcolumn} 
\restylefloat{figure} 

\makeatletter
\def\@sect#1#2#3#4#5#6[#7]#8{\ifnum #2>\c@secnumdepth
  \def\@svsec{}\else
  \refstepcounter{#1}\edef\@svsec{\csname the#1\endcsname.\hskip0.5em}\fi
  \@tempskipa #5\relax
  \ifdim \@tempskipa>\z@
    \begingroup
      #6\relax
      \@hangfrom{\hskip #3\relax\@svsec}{\interlinepenalty \@M #8\par}%
    \endgroup
    \csname #1mark\endcsname{#7}\addcontentsline
      {toc}{#1}{\ifnum #2>\c@secnumdepth \else
        \protect\numberline{\csname the#1\endcsname}\fi #7}%
  \else
    \def\@svsechd{#6\hskip #3\@svsec #8\csname #1mark\endcsname
      {#7}\addcontentsline{toc}{#1}{\ifnum #2>\c@secnumdepth \else
        \protect\numberline{\csname the#1\endcsname}\fi #7}}%
  \fi \@xsect{#5}}
\@addtoreset{equation}{section}
\makeatother
\renewcommand\thesection{\arabic{section}}
\renewcommand\theequation{\ifnum \value{section}>0
 \thesection.\arabic{equation}%
\else
\arabic{equation}%
\fi}

\newcommand{\ttbar}{t {\bar t}}
\newcommand{\mtt}{M_{t \bar t}}

\renewcommand{\thefootnote}{\small\fnsymbol{footnote}}

\begin{document}
\begin{titlepage}
  \begin{flushright}
    PITHA 08/05  
  \end{flushright}        
\vspace{0.01cm}
\begin{center}
{\LARGE {\bf Weak interaction corrections
to hadronic top quark pair production: contributions
from quark-gluon and $b \bar b$ induced reactions.}}  \\
\vspace{2cm}
{\large{\bf Werner Bernreuther\,$^{a,}$\footnote{Email:
{\tt breuther@physik.rwth-aachen.de}},
Michael F\"ucker \,$^{a,}$\footnote{Email:
{\tt fuecker@physik.rwth-aachen.de}},
 Zong-Guo Si\,$^{b,}$\footnote{Email: {\tt zgsi@sdu.edu.cn}} 
}}
\par\vspace{1cm}
$^a$Institut f\"ur Theoretische Physik, RWTH Aachen, 52056 Aachen, Germany\\
$^b$Department of Physics, Shandong University, Jinan, Shandong
250100, China\\
\par\vspace{1cm}
{\bf Abstract}\\
\parbox[t]{\textwidth}
{As an addendum to our
previous evaluation of the  weak-interaction corrections 
to hadronic  top-quark pair production \cite{Bernreuther:2006vg}
we determine the leading weak-interaction contributions 
due to the subprocesses $b {\bar b} \to  \ttbar$ and $g q \, ({\bar q})
\to \ttbar q\, ({\bar q})$.
For  several distributions in $\ttbar$ production at the LHC we find   
that  these contributions are non-negligible as compared
to the weak corrections from the other partonic subprocesses.
}
\end{center}
\vspace*{2cm}

PACS number(s): 12.15.Lk, 12.38.Bx, 13.88.+e, 14.65.Ha\\
Keywords: hadron collider physics, top quarks, QCD and electroweak 
corrections, spin effects, parity violation
\end{titlepage}
%
%
\setcounter{footnote}{0}
\renewcommand{\thefootnote}{\arabic{footnote}}
\setcounter{page}{1}
The physics of top quarks at the Tevatron and at the
upcoming Large Hadron Collider (LHC) offers the unique possibility to
explore the interactions of the heaviest known fundamental particle.
At the LHC one expects to investigate with some precision also the
so-far unknown high-energy regime, i.e., single top-quark and
top antitop-quark $(\ttbar)$ events with transverse momenta
and/or pair-invariant masses in the TeV range. The analysis and
interpretation of such events will require, in particular,
 precise standard model (SM) predictions. In this context the
electroweak corrections to hadronic $\ttbar$ production were
recently determined: the ${\cal O}(\alpha_s^2\alpha)$ 
contributions of $W, Z$ and Higgs boson
exchange to quark-antiquark annihilation $q {\bar q} \to \ttbar$ 
\cite{Bernreuther:2005is,Kuhn:2005it} and to gluon fusion
$gg \to t {\bar t}$
\cite{Bernreuther:2006vg,Kuhn:2006vh,Moretti:2006nf},
extending earlier work of\footnote{The supersymmetric QCD corrections 
to $\ttbar$ production were recently reexamined in 
\cite{Berge:2007dz}. The computation of \cite{Ross:2007ez} includes
also electroweak MSSM effects, which were analyzed  before  in
\cite{Kim:1996nza,Hollik:1997hm,Kao:1999kj}.}
\cite{Beenakker:1993yr}, and the photonic corrections to
hadronic top-quark pair production \cite{Hollik:2007sw}.

In this addendum to \cite{Bernreuther:2006vg}
we analyze  a further set of weak-interaction
corrections which we found to have some impact
on a few  kinematic distributions: i) the 
contributions of order $\alpha^2$ and $\alpha_s\alpha$ to
\begin{equation}
b{\bar b} \to t {\bar t} \, ,
\label{bbreac}
\end{equation}
and
ii) the ${\cal O}(\alpha_s\alpha^2)$ and  ${\cal O}(\alpha_s^2\alpha)$
contributions
to the reactions 
\begin{equation}
g q \; ({\bar q}) \to  t {\bar t} q\; ({\bar q}) \qquad (q=u,d,s,c,b) \, .
\label{qgreac}
\end{equation}
We employ here the so-called 5-flavor scheme \cite{Aivazis:1993pi}, where the
(anti)proton is considered to contain also $b$ and $\bar b$ quarks in
its partonic sea. Thus the reaction  (\ref{bbreac}) is a leading-order
(LO) process in this
scheme, while  (\ref{qgreac}), $q=b$,  is a next-to-leading order (NLO) QCD correction
to  (\ref{bbreac}).
The ${\cal O}(\alpha_s^2\alpha)$  corrections  to the processes
(\ref{qgreac}) were calculated already in
\cite{Bernreuther:2006vg} which we include here for completeness.
For several
top quark observables -- in particular, for the $t\bar t$
cross section -- the contributions i) and ii) are insignificant.
However, here we show
that for the  pair-invariant mass distribution and for 
the top-quark
helicity asymmetry, which are among the key observables in the tool-kit  
for  search of new physics in $t \bar t$ events, these  corrections 
do matter if one aims  at predictions with a precision
at the percent level.

The amplitude of  (\ref{bbreac}) receives, in  Born approximation and putting
$m_b=0$, the following
contributions: 
a) $t$-channel $W$ boson exchange
$b {\bar b} {\stackrel{ W}{\longrightarrow}} t {\bar t}$,
b) $s$-channel photon and  $Z$ boson exchanges $b {\bar b}
{\stackrel{\gamma, Z}{\longrightarrow}} t {\bar t} $
and c) $s$-channel gluon exchange $b {\bar b} {\stackrel{
    g}{\longrightarrow}} t {\bar t}$.
The  $t$-channel $W$ boson exchange contribution a) is
not suppressed by a small
Cabibbo-Kobayashi-Maskawa (CKM) mixing matrix element, in contrast to the
corresponding $t$-channel amplitudes 
$b {\bar d}, b {\bar s} \to t \bar t$ + c.c. channels.

The lowest order  weak-interaction induced contribution to the squared
transition matrix element  $|{\cal M}(b {\bar b} \to t \bar t)|^2$
are of order $\alpha^2$ and $\alpha_s\alpha$; the latter arises from
the interference of the amplitudes a) and c). 

The dominant part of  $|{\cal M}(b {\bar b} \to t \bar t)|^2$
is due to $W$ exchange a), as can be understood from inspecting the
various terms  in the limit of large parton center-of-mass
energy $\sqrt{\hat s} \gg 2 m_t$. It has the following properties:
First, it is positive while 
the  weak-interaction corrections to
 $gg, q {\bar q} \to t \bar t$ $(q \neq b)$ are negative in most of the
kinematic range of $\hat s$. 
Second, $t$-channel $W$ exchange produces top quarks mostly in the
forward region. Thus one expects these contributions to be relevant
only for relatively small transverse momentum $p_T$ of the
(anti)top-quark. For the distribution of the pair-invariant mass 
 $M_{t \bar t}=\sqrt{(p_t +p_{\bar t})^2}$ no such conclusion can be
 drawn. Third, $t$-channel $W$ exchange generates left-handed top
 quarks and right-handed antitop-quarks. In the high-energy regime
$M_{t \bar t} \gg m_t$, where the top quarks behave more and more like
massless quarks, the (anti)top quarks due to a) have, therefore,
(positive) negative helicity. 

\begin{figure}[H]
\begin{center}
\includegraphics[width=14cm]{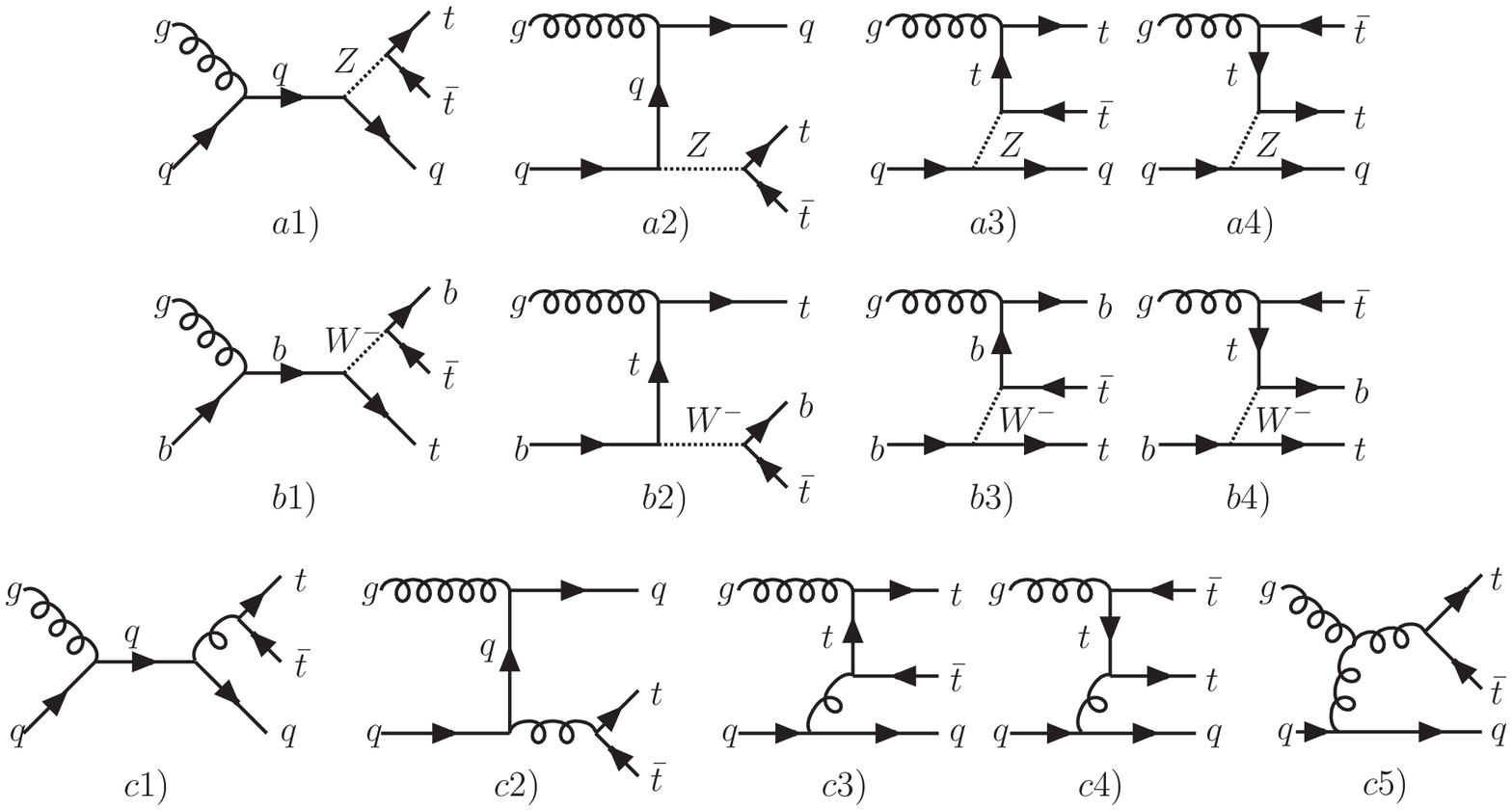}
\end{center}
\caption{Feynman diagrams 
for $g q  \, ({\bar q}) \to  \ttbar  q \, ({\bar q})$
 to leading order in the weak and strong interactions.}
 \label{fig:feynd}
 \end{figure}

The Feynman diagrams for the reactions (\ref{qgreac})
 are shown in Fig.~\ref{fig:feynd} to
leading order in the weak and strong interactions. 
The exchange  of the SM Higgs boson is numerically insignificant
and therefore not taken into account. The  diagrams
Fig.~\ref{fig:feynd}b1 - Fig.~\ref{fig:feynd}b4
with  $W$-boson exchange  are
relevant only for $b$ quarks in the initial state. For $q=s,d$ the
corresponding amplitudes are suppressed by small
CKM mixing 
matrix elements  ($|V_{td}|\sim 7 \times
10^{-3}$ and $|V_{ts}|\sim 3.5 \times 10^{-2}$ \cite{Yao:2006px}).
Here we compute the 
${\cal O}(\alpha_s\alpha^2)$ contributions to the squared matrix
elements (Figs.~\ref{fig:feynd}a for $q\neq b$ and
Figs.~\ref{fig:feynd}a
and~\ref{fig:feynd}b for $q=b$). The terms corresponding to the squares
 of Figs.~\ref{fig:feynd}a2 and~\ref{fig:feynd}b3,
 their interference, and the interference between Fig.~\ref{fig:feynd}b3 and
 Fig.~\ref{fig:feynd}c2
 have initial-state collinear singularities
which we removed within the standard ${\overline{\rm MS}}$
factorization scheme. These terms are therefore expected to exhibt some
sensitivity to variations of the factorization scale $\mu_F$. For
completeness, we take into account in the numerical evaluations below
also the weak-interaction corrections of ${\cal O}(\alpha_s^2\alpha)$ (i.e., the
interferences Figs.~\ref{fig:feynd}a and~\ref{fig:feynd}c for $q\neq
b$ and  Figs.~\ref{fig:feynd}a,b and~\ref{fig:feynd}c for $q = b$) which
were computed in \cite{Bernreuther:2006vg}. 

The qualitative discussion in the previous paragraphs
 is corrobated by
the numerical evaluation of  the corrections i) and ii). 
As far as the contributions of these
terms to the hadronic $t\bar t$ production cross sections 
at the Tevatron and at the LHC are concerned, they are below the
percent level and are, like the electroweak contributions
from the other partonic 
subprocesses \cite{Bernreuther:2006vg,Kuhn:2005it,Hollik:2007sw}, smaller than the
uncertainties of the present QCD predictions.
Next we analyze three distributions relevant for top physics: the
transverse momentum distribution, the pair-invariant mass
distribution, and the helicity asymmetry. At the Tevatron (i.e., for $p{\bar p}$
collisions at $\sqrt{s}=$1.96 TeV) $b$-quark induced $t \bar t$
production plays no role, and the weak-interaction
induced contributions to these distributions from (\ref{bbreac}) and
(\ref{qgreac}) are completely negligible. However, they matter
for the LHC, i.e., 
for $p \, p \, \to \, t {\bar t}\,X$  at $\sqrt{s}=$14 TeV. 

Let us now discuss the $p_T$ and $\mtt$  distribution and the top-quark
helicity asymmetry for the LHC.   We compare the corrections i) and
ii) with the weak-interaction induced contributions of order
$\alpha^2$ and $\alpha_s^2\alpha$  due to the subprocesses
$q {\bar q} \to t {\bar t}$ $(q\neq b)$ and $gg  \to t {\bar
  t}$ \cite{Bernreuther:2005is,Bernreuther:2006vg}, which we denote
 by corrections  iii) in the following.
These depend on the unknown SM Higgs boson mass, for which we
use the values $m_H=$ 120 GeV and 200 GeV. The corrections 
i), ii), and iii) will be  normalized to the respective distributions
 $d\sigma_{LO}$  obtained in lowest-order QCD 
 from $q{\bar q}, gg \to \ttbar$. In the case of the parity-violating
helicity asymmetry, i), ii), and iii) are normalized to $d\sigma_{LO}/d\mtt$.
As in \cite{Bernreuther:2006vg} we use $m_t=172.7$ GeV,
$\alpha_s(2m_t)=0.1$, and  $\alpha(2m_t)=1/126.3$.
The LO QCD terms and the contributions of i) and iii) to the distributions
are evaluated with the  LO parton distribution functions (PDF)
CTEQ6.L1, while for the computation
 of the contributions from ii), which depend on the factorization
scale, the set CTEQ6.1M  \cite{Pumplin:2002vw} is used. 
The scale $\mu_F$ is varied between 
$m_t/2\leq\mu_F\leq 2 m_t$.  Dependence on the renormalization scale $\mu_R$
enters only via the ${\overline{\rm MS}}$ coupling $\alpha_s$. The ratio
of the corrections iii) and $d\sigma_{LO}$ is practically independent
of $\alpha_s$, while the corresponding ratios involving i) and ii)
vary weakly with $\mu_R$.

\begin{figure}[H]
\begin{center}
\includegraphics[width=7cm, height=7cm]{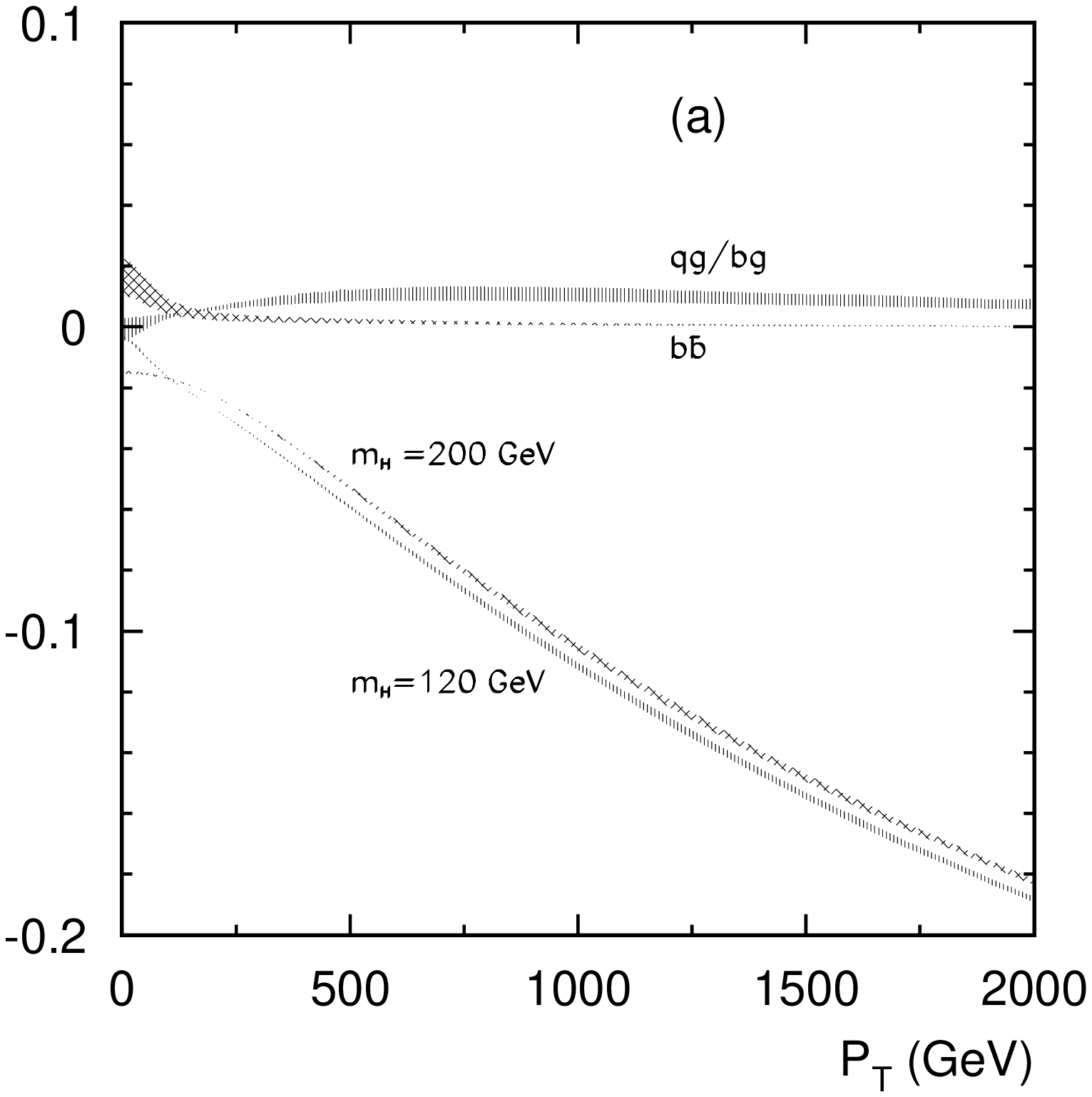}
\includegraphics[width=7cm, height=7cm]{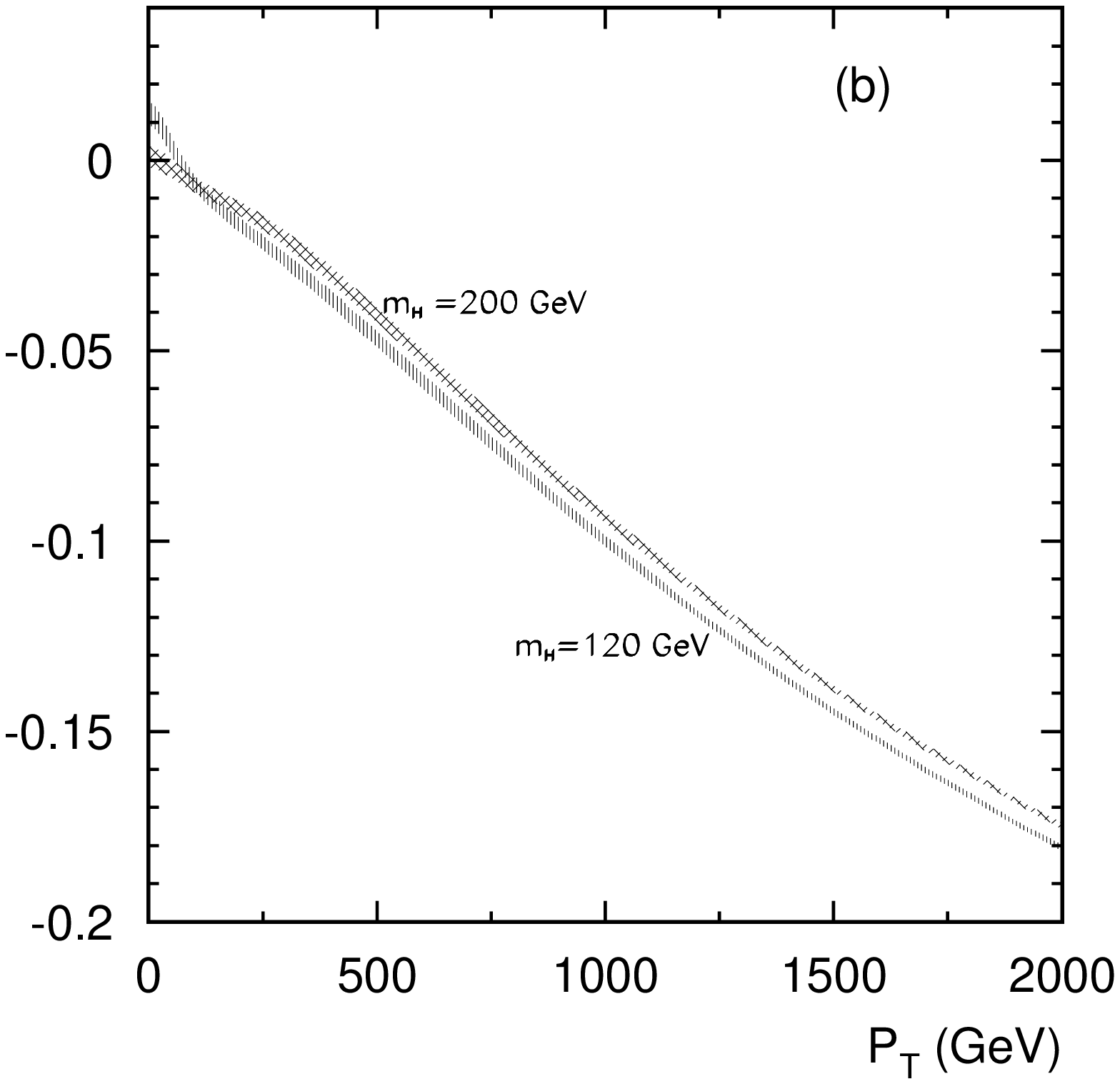}
\end{center}
\caption{ a) Ratios  
$(d\sigma_{weak}/dp_T)/(d\sigma_{LO}/ dp_T)$ where $d\sigma_{weak}$  are
the weak-interaction corrections i), ii), and iii)
to the reactions (\ref{bbreac}), (\ref{qgreac}),
 and $q {\bar q}, gg \to \ttbar$ $(q\neq b)$, respectively. The latter
 corrections are shown for two different values of the Higgs boson
 mass. The hatched areas arise from scale variations as described in
 the text.  b) Sum of the ratios shown in a) for 
two different values of $m_H$.}
 \label{fig:dpt}
 \end{figure}

Fig.~\ref{fig:dpt}a shows the various 
weak-interaction contributions to the transverse
momentum distribution of the top quark at the LHC, normalized
to $d\sigma_{LO}/dp_T$. The hatched areas depict the range of values
when $\mu\equiv\mu_F=\mu_R$ is varied between $m_t/2$ and $2m_t$.
Fig.~\ref{fig:dpt}a  shows that the weak correction
 i) to the $p_T$ distribution of the top quark is positive, as
 expected, and small. 
Its significance is confined to the region $p_T \lesssim$ 100 GeV,
where it dominates the
other weak corrections. However, in this region
these corrections make up only 
 between 1$\%$ and 2$\%$ of the  LO QCD $p_T$ distribution.
In the high $p_T$ region, where the weak-interaction corrections to
the $p_T$ spectrum become relevant, the  contribution
from the processes (\ref{bbreac}) and (\ref{qgreac})
do not matter in comparison to the  weak
corrections iii).
Fig.~\ref{fig:dpt}b  displays the ratio of the sum of the weak corrections
i), ii), and iii) and the LO QCD contribution.

\begin{figure}[H]
\begin{center}
\includegraphics[width=7cm, height=7cm]{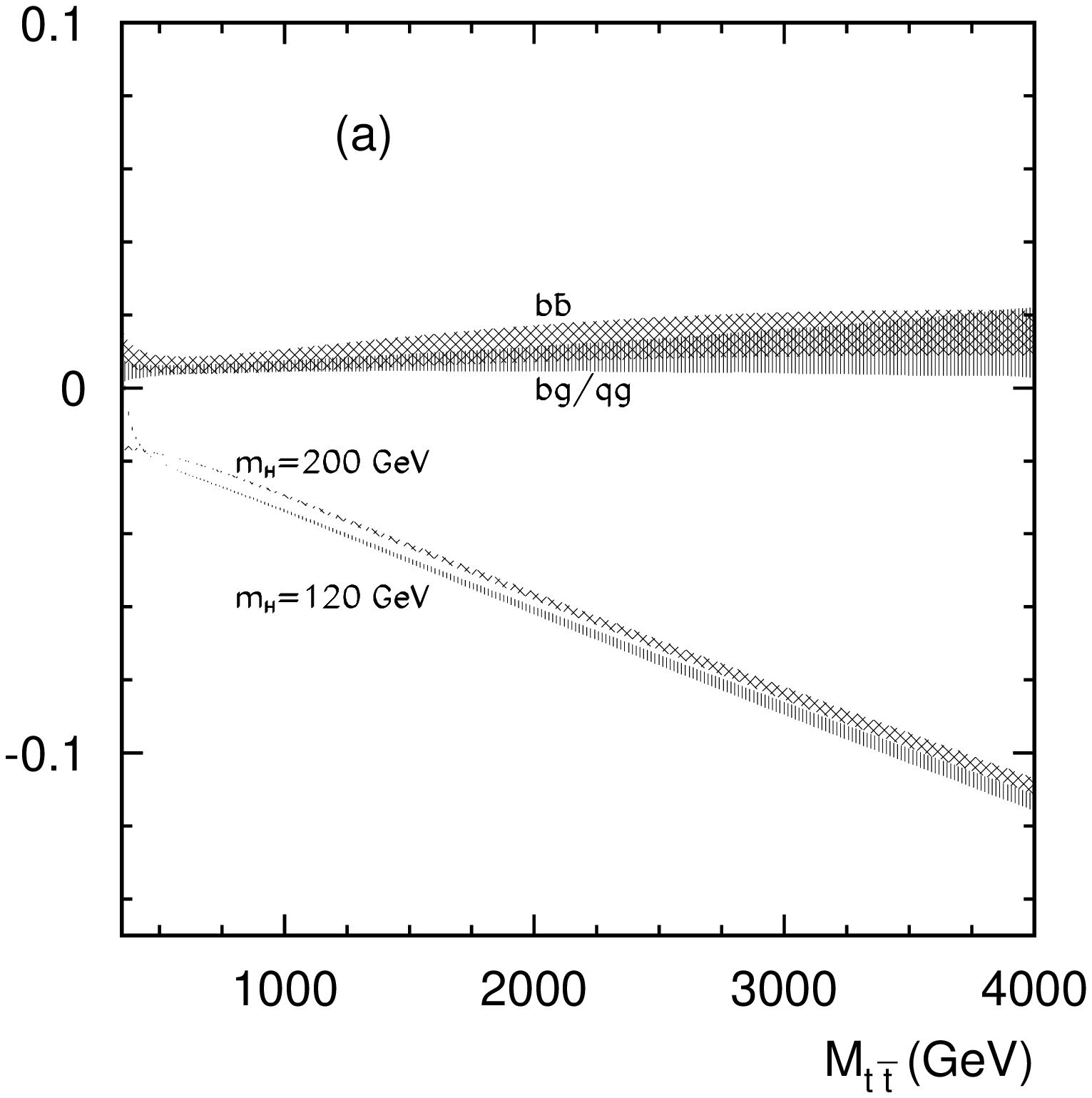}
\includegraphics[width=7cm, height=7cm]{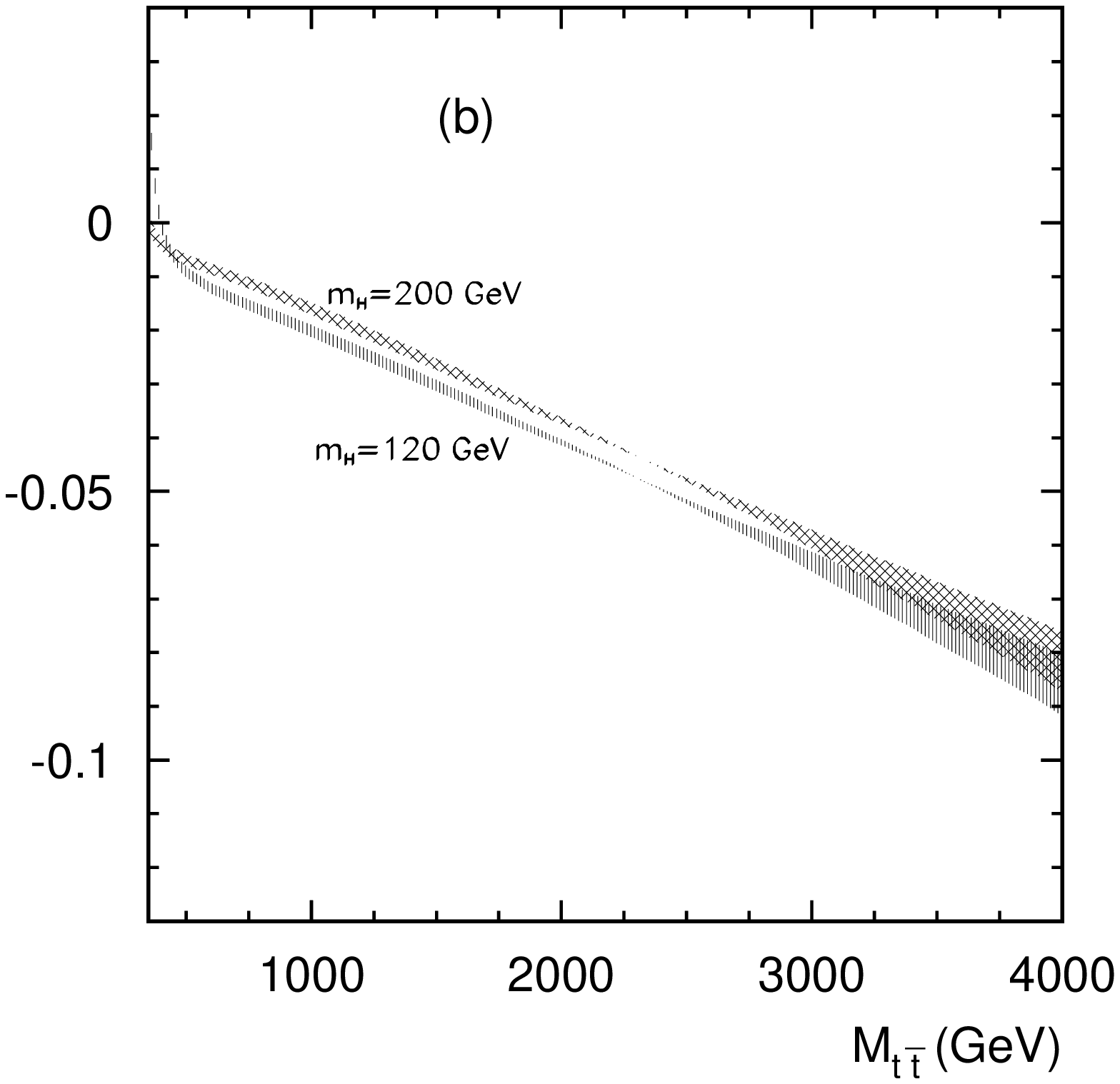}
\end{center}
\caption{ a) Ratios  $(d\sigma_{weak}/d\mtt)/(d\sigma_{LO}/d\mtt)$ 
where $d\sigma_{weak}$  refers to
the weak-interaction corrections i), ii), and iii).
 The latter
 corrections are shown for two different values of the Higgs boson
 mass. The hatched areas arise from scale variations as described in
 the text.  b) Sum of the ratios shown in a)
for two different values of $m_H$.
}\label{fig:dmtt}
 \end{figure}

In Fig.~\ref{fig:dmtt}a the analogous ratios are displayed for the 
$M_{t \bar t}$ distribution. The weak-interaction corrections
i) and ii) are both positive and show a considerable scale uncertainty.
They reduce the magnitude of the leading weak corrections iii), which
are negative, by an amount of about $50\%$, as shown in
Fig.~\ref{fig:dmtt}b.

\begin{figure}[H]
\begin{center}
\includegraphics[width=7cm, height=7cm]{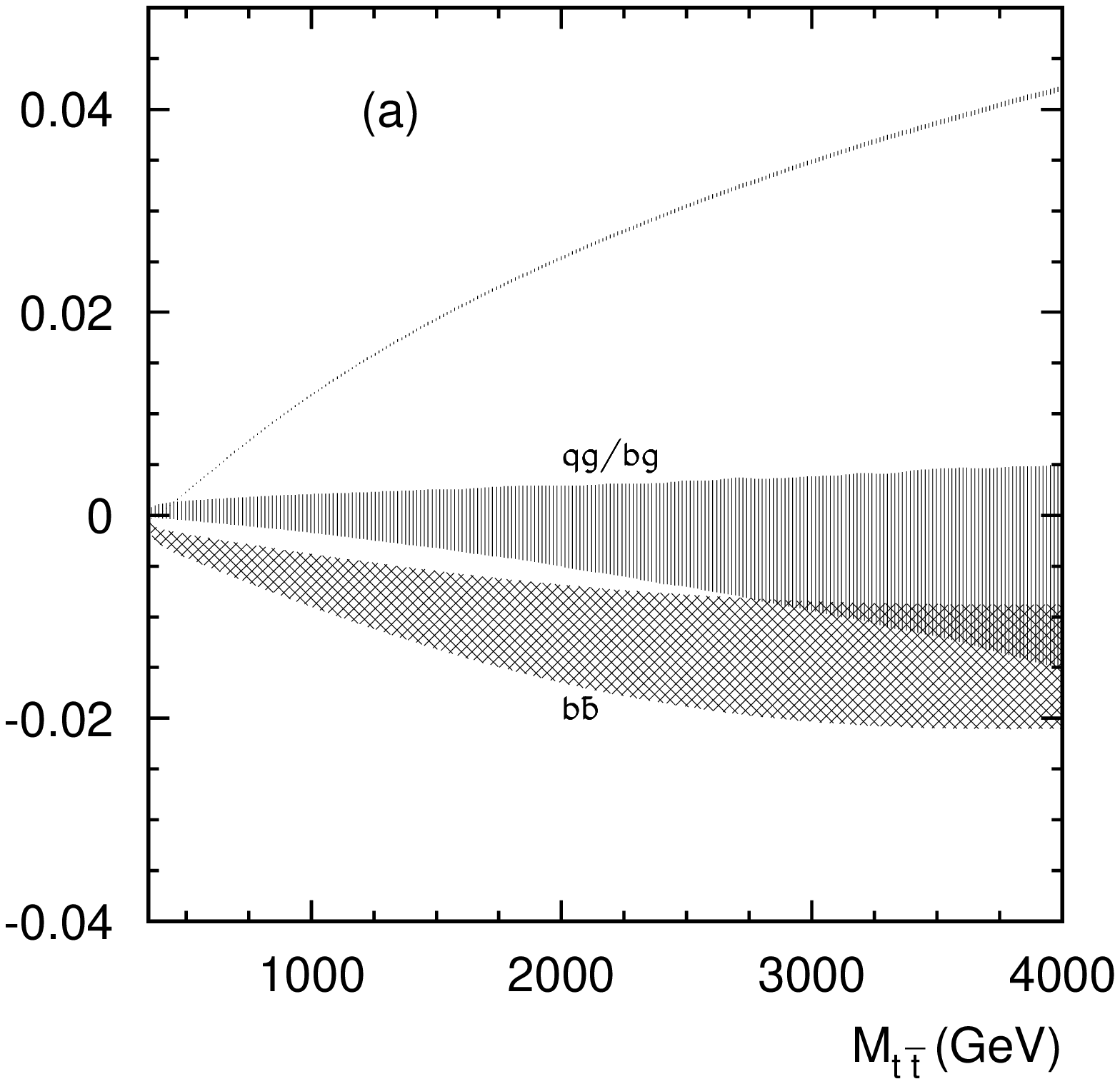}
\includegraphics[width=7cm, height=7cm]{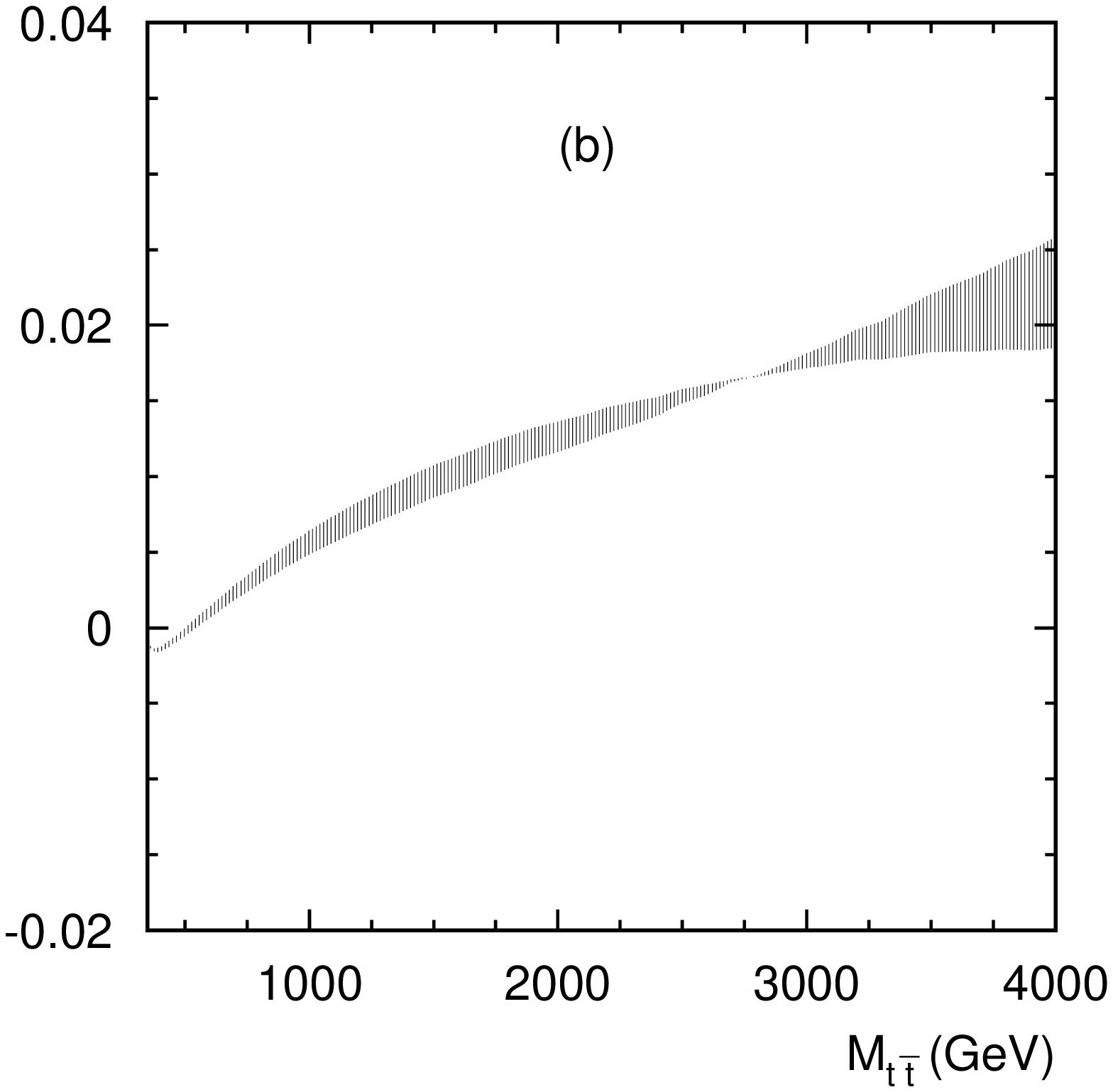}
\end{center}
\caption{
a) Contribution of the various partonic subprocesses to the helicity
asymmetry $\label{helas}$:  initial states 
$q \bar q$ $(q\neq b)$ and $gg$ (thin line), $qg$ and ${\bar q}g$
$(q=u,...,b)$ (vertically hatched area), and $b \bar b$ (cross hatched
 area).  b) Sum of the three
contributions shown in a).
}\label{fig:dheli}
 \end{figure}

Finally, we consider the parity-violating
helicity asymmetry for $t$ quarks defined by
\begin{equation}
\Delta_{hel}\, = \, \frac{Z_{hel}}{d\sigma_{LO}/d\mtt}\, , \qquad 
Z_{hel}\, = \,  \frac{d\sigma_{+}}{d\mtt} - 
\frac{d\sigma_{-} }{d\mtt}  \, .
\label{helasy}
\end{equation}
The subscripts $\pm$ in (\ref{helasy}) refer to a  $t$ quark with
positive/negative
helicity while the helicity states of the $\bar t$ are summed. 
(In  \cite{Bernreuther:2006vg} a different normalization was chosen for 
$\Delta_{hel}$.) Fig.~\ref{fig:dheli}a displays  the weak-interaction 
induced contributions  i), ii) and iii) 
to $\Delta_{hel}$. (As the SM Yukawa coupling is parity-conserving,
iii) does not depend on $m_H$.) Each correction i) and ii) 
 shows a considerable
scale dependence which, however, cancels to a large extent in the sum
of the two contributions -- c.f. Fig.~\ref{fig:dheli}b.
The corrections i) and ii) reduce the contribution iii) to
$\Delta_{hel}$ by about $50\%$. The $t$ quark helicity asymmetry 
in the SM is then $\Delta_{hel} \lesssim 2\%$  for $\mtt \lesssim$ 4 TeV.
Such a small effect will hardly be measurable at the
LHC. Nevertheless, as emphasized in \cite{Bernreuther:2006vg}, 
this observable is an
ideal experimental 
sensor for tracing possible new parity-violating interactions
in $t \bar t$ production; thus $\Delta_{hel}$ should be computed as
precisely as possible within the SM.

If one takes into account  only
$t \bar t$ events with  $p_T \geq p_{Tmin}$, the corrections
i), ii)  will not change significantly, as long as  $p_{Tmin}$
is not too large. Choosing, for
instance, $p_{Tmin}=$ 30 GeV does not lead to a significant change 
of the results shown in Figs.~\ref{fig:dpt}~-~\ref{fig:dheli}.  
Eventually, the weak
corrections to these distributions discussed here 
should be evaluated  \cite{bfs07} in conjunction 
with the known NLO QCD
corrections, for which NLO PDF, in particular a
 NLO $b$-quark PDF is to be used. 
(For recent updates of this PDF,
see \cite{Tung:2006tb,Martin:2007bv}). The NLO  $b$-quark
PDF enhances the $b$-quark induced
weak contribution to the $\mtt$ distribution and to $\Delta_{hel}$ at
large $\mtt$.

In conclusion we have determined for
hadronic $t \bar t$ production the leading weak-interaction corrections
due to the subprocesses  $b {\bar b} \to t {\bar t}$
and  $g q \, ({\bar q})
\to \ttbar q\, ({\bar q})$.
For the LHC we find that 
in the case of the pair-invariant mass distribution and of the
helicity asymmetry these contributions are non-negligible as compared
to the weak corrections from $q {\bar q}, gg \to \ttbar$.
As these distributions are key observables for investigating the
interactions of top quarks in the high-energy regime these corrections
should be taken into account when it comes to precision analyses
 of  future $t \bar
t$ events at the LHC.

\subsubsection*{Acknowledgments}
We thank P. Uwer for discussions.
Z.G. Si wishes to thank the Physics 
Department of RWTH Aachen, where  part of this
work was done, for its hospitality and the
A. v. Humboldt Stiftung for financial support. 
This work was also supported by Deutsche Forschungsgemeinschaft 
SFB/TR9.


\newpage

%
%
%
%

\end{document}